# Biocompatible surface functionalization architecture for a diamond quantum sensor

Mouzhe Xie^1, Xiaofei Yu^2, Lila V. H. Rodgers[3], Daohong Xu[2], Ignacio Chi-Durán[4], Adrien Toros[5], Niels Quack[6], Nathalie P. de Leon[3], Peter C. Maurer[1]

[1] Pritzker School of Molecular Engineering, the University of Chicago, Chicago, IL 60637, USA
[2] Department of Physics, the University of Chicago, Chicago, IL 60637, USA
[3] Department of Electrical and Computer Engineering, Princeton University, Princeton, NJ 08544, USA
[4] Department of Chemistry, the University of Chicago, Chicago, IL 60637, USA
[5] Center of MicroNanoTechnology, École polytechnique fédérale de Lausanne, Lausanne, CH-1015, Switzerland
[6] Institute of Microengineering, École polytechnique fédérale de Lausanne, Lausanne, CH-1015, Switzerland
^ These authors contributed equally to this work.
Correspondence and requests for materials should be addressed to pmaurer@uchicago.edu

**Quantum metrology enables some of the most precise measurements. In the life sciences, diamond-based quantum sensing has enabled a new class of biophysical sensors and diagnostic devices that are being investigated as a platform for cancer screening and ultra-sensitive immunoassays. However, a broader application in the life sciences based on nanoscale nuclear magnetic resonance spectroscopy has been hampered by the need to interface highly sensitive quantum bit (qubit) sensors with their biological targets. Here, we demonstrate a new approach that combines quantum engineering with single-molecule biophysics to immobilize individual proteins and DNA molecules on the surface of a bulk diamond crystal that hosts coherent nitrogen vacancy qubit sensors. Our thin (sub-5 nm) functionalization architecture provides precise control over protein adsorption density and results in near-surface qubit coherence approaching 100 μs. The developed architecture remains chemically stable under physiological conditions for over five days, making our technique compatible with most biophysical and biomedical applications.**

Recent developments in quantum engineering and diamond processing have brought us considerably closer to performing nanoscale nuclear magnetic resonance (NMR) and electron paramagnetic resonance (EPR) spectroscopy of small ensembles and even individual biomolecules. Notably, these advances have enabled the detection of the nuclear spin noise from a single ubiquitin protein (1), and the probing of the EPR spectrum of an individual paramagnetic spin label conjugated to a protein (2) or DNA molecule (3). More recently, lock-in detection and signal reconstruction techniques (4, 5) have enabled one- and multidimensional NMR spectroscopy with 0.5 Hz spectral resolution (6–8). More advanced control sequences at cryogenic temperatures have further enabled mapping the precise location of up to 27 $^{13}$C nuclear spins inside of diamond (9). Yet, biologically meaningful spectroscopy on intact biomolecules remains elusive. One of the main outstanding challenges, which is required to perform nanoscale magnetic resonance spectroscopy of biomolecules, is the need to immobilize the target molecules within the 10-30 nm sensing range (2, 3, 7) of a highly coherent nitrogen vacancy (NV) qubit sensor. Immobilization is necessary because an untethered molecule would otherwise diffuse out of the detection volume within a few tens of microseconds.

Various avenues to diamond functionalization have been pursued over the last decade (10–12). However, none of the currently known approaches has led to the desired results of interfacing a coherent quantum sensor with target biomolecules. For example, hydrogen terminated diamond surfaces can be biochemically modified and form biostable surfaces (10, 13); but near surface NV centers are generally charge-unstable under hydrogen termination (14), posing open challenges for NV sensing.  On the other hand, oxygen terminated diamond surfaces have been used to create charge stable NV$^-$ centers with exceptional coherence times within 10 nm from the diamond surface (15). However, perfectly oxygen-terminated diamond surfaces generally lack chemically functionalizable surface groups (such as carboxyl or hydroxyl groups), making it difficult to control immobilization density and surface passivation. Our approach (Fig. 1*A*) overcomes these limitations by utilizing a 2 nm-thick $Al_2O_3$ layer deposited onto an oxygen terminated diamond surface by atomic layer deposition (ALD). This $Al_2O_3$ "adhesion" layer is silanized by *N*-[3-(trimethoxysilyl)propyl]ethylenediamine to create an amine (–$NH_2$) terminated surface, which in turn is then grafted with a monolayer of heterobifunctional polyethylene glycol (PEG) via a *N*-hydroxysuccinimide (NHS) reaction, a process also referred as PEGylation. The PEG layer serves two purposes. First, it passivates the diamond surface to prevent nonspecific adsorption of biomolecules. Second, by adjusting the density of PEG molecules with functional groups (e.g., biotin or azide), we can control the immobilization density of proteins or DNA target molecules on the diamond surface. Furthermore, the small persistence length of the PEG linker (approximately 0.35 nm) allows the immobilized biomolecules to undergo rotational diffusion (16). This tumbling motion is the basis for motional averaging of the NMR spectra and helps to prevent immobilization of molecules in biologically inactive orientations.

Diamond-based sensing critically relies on minimizing the thickness of any functionalization layer while maintaining excellent surface morphology and surface coverage. We hence carefully characterized the surface at each step of our functionalization procedure. As confirmed by atomic force microscopy (AFM), thermal ALD enabled us to deposit a uniform 2 nm-thick $Al_2O_3$ layer of excellent surface morphology (arithmetical mean deviation $R_a$ = 459 pm) on an oxygen-terminated diamond surface ($R_a$ = 446 pm) (Fig. 1 *B* and *F*). The changes in surface properties are corroborated by contact angle measurement (see Fig. S7 in the Supporting Information, abbreviated as SI thereafter). A slight increase of surface roughness can be observed after treatment with 10 mM KOH for 10 s, which serves the purpose of -OH activation for silanization ($R_a$ = 841 pm) but also leads to hydrolysis of $Al_2O_3$. A final surface roughness of $R_a$ = 866 pm can be observed after PEGylation. X-ray photoelectron spectroscopy (XPS) further confirmed the presence of aluminum (Al2p) after each surface treatment step, indicating that the $Al_2O_3$ layer remains stable during the processing (Fig. 1*C*). Angle-resolved XPS (ARXPS) allowed us to further estimate the thickness of the $Al_2O_3$ and PEG layer to be 2.0 ± 0.1 nm and 1.2 ± 0.2 nm (see Supplementary Note 2 in SI). We note that ARXPS likely underestimated the true thickness of the PEG layer since ARXPS was performed under ultrahigh vacuum, which lead to a collapse of the PEG layer. Assuming a Gaussian chain model, we can estimate the thickness of the hydrated PEG layer to be 2.8 nm (see Supplementary Note 2 in SI). From this we estimate that the total thickness of the functionalization layer is on the order of 5 nm. Notably, shorter PEG can be employed to further reduce the overall thickness to 3 nm without impeding the fine control over the grafting density (*vide infra*), as demonstrated in the Supplementary Figure S8.

Next, we turned our attention to controlling and characterizing the adsorption density of proteins on a diamond surface. The density of binding sites can be controlled by adjusting the stoichiometric ratio of methyl-terminated PEG (mPEG) and functional PEG groups, for example, biotin-terminated PEG (biotinPEG) or azide-terminated PEG (azidePEG) for click chemistry (Fig. 1*E*). To investigate the effectiveness of our method, we characterized the adsorption density of Alexa-488-dye-labeled streptavidin (SA-488), our model system, by single-molecule fluorescence microscopy (17). Figure 2*A* shows a series of epiluminescence images for diamond samples with varying biotinPEG density that were incubated in 7 nM SA-488 for 20 min. The number of fluorescent spots, i.e. individual SA molecules, shows a clear dependence of the adsorption density on the biotinPEG percentage. Importantly, for the diamonds coated solely with mPEG, we observed 4×10$^{-3}$ SA-488 protein per µm$^2$, whereas for 2% biotin we found roughly 0.5 SA-488 per µm$^2$ (Fig. S9). This suggests that we can control the protein adsorption density over more than two orders of magnitude. We note that for higher biotinPEG densities individual SA-488 molecules were no longer optically resolvable. Furthermore, the immobilization of SA-488 was spatially homogeneous and highly reproducible, in sharp contrast to control experiments that skipped the silanization step, which resulted in a heterogeneous distribution of SA-488 on the diamond surface (Fig. S10). Performing these titration series for two different $Al_2O_3$ layer thickness, we observed no qualitative difference in SA-488 adsorption density between a 2 nm and a 50 nm $Al_2O_3$ layer, indicating that working with an ultrathin $Al_2O_3$ layer does not negatively impact bio-functionalization. Interestingly, we did observe a reduction in the signal-to-noise ratio in our fluorescent microscopy images. At least partially, this observation can be explained by self-interference of an emitter at the diamond-$Al_2O_3$ interface (see Supplementary Note 3 in SI).

The developed diamond surface modification architecture can be readily combined with most well-established biochemical conjugation techniques. We demonstrate the versatility of our approach with two examples of bioconjugation: first, a molecular biological conjugation of target molecules to the diamond surface via a biotin-streptavidin interaction and, second, a biochemical conjugation via "click chemistry". For both examples, a Cy3-labeled 40-nt single stranded DNA (Cy3-ssDNA) served as a model molecule. In the first system (Fig. 2*B*), biotinylated Cy3-ssDNA was immobilized to the diamond surface mediated by streptavidin with no fluorescent label (nSA). Time-dependent fluorescence measurements show a single-step decay in the fluorescent signal for the majority of fluorescent spots, a hallmark for single-molecule measurements (Fig. 2*D*) (18). In addition, we also observed 2-step decay events, which can be explained by multiple ssDNA molecules binding to a single nSA homotetramer or coincidental co-location of two nSA molecules. The second system exploited strain-promoted azide-alkyne cycloaddition (SPAAC), also known as "copper-free click chemistry" (Fig. 1*E*), for its reliable performance and fast kinetics (19). The diamond surfaces were prepared following the same procedure except that biotinPEG was replaced by an azidePEG compound. Through SPAAC, the same Cy3-ssDNA engineered with a dibenzocyclooctyne (DBCO)-label at its 5' was successfully immobilized to diamond surfaces (Figs. 2*C* and S11). The large fluorescence spots, which include more than one fluorophore, were likely originated from the presence of aggregates in the DBCO-tagged Cy3-ssDNA sample.

For any practical biophysical or diagnostics applications, it is important that the functionalization layer maintains good chemical stability without degrading over the course of a typical experiment. Oxides are known to hydrolyze over time when exposed to water, at a rate that depends on the film (e.g., composition, deposition method, and film quality) and solvent (e.g., pH and salinity) properties (20). Because the exact mechanism is difficult to predict, we experimentally tested the chemical stability of our functionalization architecture. We immobilized SA-488 to the diamond surface and monitored their fluorescence over the course of a week while storing the sample in a sodium phosphate buffer (pH 7.4, [$NaH_2PO_4$ + $Na_2HPO_4$] = 50 mM, [NaCl] = 100 mM) at 23.5°C room temperature. Figure 3*A* shows the observed fluorescent signal over time. The decrease in number of SA-488 per field-of-view can be attributed to either a dissociation of the functionalization layer or photobleaching of the SA-488. This sets an upper limit for the functional layer dissociation rate to a half-life time of 5.7 days. In addition to these optical measurements, we also used AFM to monitor changes of the $Al_2O_3$-layer thickness as a function of submersion time in doubly deionized (MilliQ) water and sodium phosphate buffer. Figure 3*B* shows the thickness of a lithographically patterned $Al_2O_3$ structure as a function of submersion time measured by AFM. We note that in water the dissociation is negligible (observed rate 0.19 ± 0.54 nm/day with a fairly large uncertainty), which is in good agreement with the optical measurements. However, for the measurements in sodium phosphate buffer, we determined a dissociation rate of 0.74 ± 0.22 nm/day, which is slightly larger than what would be expected from the optical measurements in Figure 3*A* and from the direct optical observation of the lithographic $Al_2O_3$ patterns in Figure S12.

In parallel, we studied the impact of our functionalization architecture on the spin coherence ($T_2$) of near-surface NV centers. Long coherence times are essential to NV-based quantum sensing because the sensitivity is generally proportional to $\sim \sqrt{T_2}$ (21). Figure 4*B* shows an example of the coherence time of a NV center under (YY-8)$_{N=8}$ dynamical decoupling before and after surface modification (see Fig. S15 for pulse sequence diagram). YY-8 sequences were chosen for their robustness to pulse errors and the ability to suppress spurious signals from nearby nuclear spins (22). The observed coherence follows a stretched exponential $\exp[-(t/T_2)^n]$, with $T_2$ = 47 µs before and $T_2$ = 31 µs after the functionalization (the exponent *n* in Fig. 4 *B* and *C* ranges from 1 to 1.8). We further systematically investigated $T_2$ and the longitudinal spin relaxation ($T_1$) times of 8 spatially resolved NV centers with depth ranging from 2.3 nm to 11 nm (Fig. 4*C*, and for NV position Fig. 4*A*), where the NV depths were determined by probing noise from the environmental $^1$H spins in the immersion oil following the method described in (23). All investigated NV centers, with the exception of the shallowest NV (depth 2.3 nm), maintained their coherence after functionalization, with an observed characteristic increase in $T_2$ as a function of NV depth (15). Overall, the $T_2$ of these NVs upon surface functionalization is reduced by 49 ± 22% under (YY-8)$_{N=8}$, or 15 ± 18% when using spin-echo sequence (Fig. S16). A careful investigation on spectral decomposition (see SI for method) reveals a broadband noise spectrum across the frequency range of 0.05-10 MHz (Figs. 4*B* and S17). We did not register a sizable reduction in $T_1$ time (Fig. S18) after surface treatment, suggesting that charge and magnetic field noise spectra have neglectable frequency components at 3 GHz.

In conclusion, we developed a chemically stable universal diamond surface functionalization architecture that can be combined with most of the established biochemical conjugation techniques. While we demonstrated biotin-SA conjugation and SPAAC click chemistry, our functionalization approach can be readily extended to other conjugation techniques, such as maleimide reaction (24), $Ni^{2+}$/His-tag interaction (25), and sortase-mediated enzymatic conjugation (26). Combined with single-molecule fluorescence imaging techniques, we have shown that this architecture allows for a precise control over the conjugation density of individual target proteins and DNA molecules. The observed NV coherence times of up to 100 µs are long enough to perform highly sensitive state-of-the-art quantum sensing experiments on biological targets (6–8). Based on the demonstrated sensor-target distances and qubit coherence, we predict that the NMR signal of an individual $^{13}$C nuclear spin can be detected with integration times as short as 100 seconds (see Methods). The anticipated integration time can further be reduced by minimizing the overall thickness of the functionalization layer and increasing the NV coherence time. A decrease in the functionalization layer thickness can be achieved by the deposition of sub-1 nm $Al_2O_3$ layer and the passivation with shorter PEG, whereas the coherence time can be increased through further material processing, such as optimization of $Al_2O_3$ growth parameters and additional annealing after $Al_2O_3$ deposition, as well as an increasing the number of $\pi$-pulses during dynamical decoupling.

During the preparation of this manuscript, we became aware of recent work described in Ref. (27), which applied NV sensing to probe the NMR signature of a self-assembled monolayer of organic molecules on an $Al_2O_3$-coated diamond sensors. Combining this (27) and other NV sensing techniques (6–8) with our molecular "pulldown" experiments will enable NMR and EPR spectroscopy of intact biomolecules in a relevant biological environment. Existing microfluidics platforms (28) can readily be combined with our diamond passivation and functionalization method, which will pave the way to label-free high-throughput biosensing with applications in quality management in the pharmaceutical industry (29, 30), single-cell screening for metabolomics (31), proteomics (32), and screening of cancer markers (33). Furthermore, positioning individual biomolecules within the 10 nm sensing range of a single NV center brings us closer to performing EPR and NMR spectroscopy on individual intact biomolecules. Such a technology would enable the *in vivo* detection of receptor-ligand binding (such as pharmaceuticals or toxins) events, post-translational protein modification (such as phosphorylation), and small conformational changes that might be difficult to detect by existing methods.

## Materials and Methods

### Functionalization

Single crystalline diamonds slabs (2×2×0.5 mm$^3$, ElementSix, electronic grade, Cat# 145-500-0385) were sonicated in acetone and isopropanol for 5 min each and dried with nitrogen gas before $Al_2O_3$ deposition. The deposition of $Al_2O_3$ layer was carried out in a Ultratech/Cambridge Savannah ALD System by alternatingly delivering trimethylaluminum and $H_2O$ gases at 200°C, 20 cycles for 2 nm layer and 550 cycles for 50 nm $Al_2O_3$ thickness. The diamonds were then soaked in 100 mM KOH for 10 s before being rinsed with a copious amount of water, and dried in an oven set to 80°C. Silanization was achieved using freshly prepared 3% *N*-[3-(trimethoxysilyl)propyl]ethylenediamine solution in anhydrous acetone at room temperature for 20 min. Upon completion, the surfaces were rinsed with acetone and water, and dried with nitrogen gas. For PEGylation, solutions of heterobifunctional PEG with various molecular weights were freshly prepared at ~0.5 M concentration in 5 mM $NaHCO_3$ buffer with a final pH between 8.0-8.5. Finally, the diamonds were immersed in the PEG solutions and incubated for 1 to 2 hours in dark at room temperature, before being extensively washed with water and dried with nitrogen gas.

### Immobilization of biomolecules

To immobilized biomolecules, 3 uL of 7 nM Alexa-488 labeled streptavidin (in 50 mM pH 7.4 sodium phosphate buffer that also contained 100 mM NaCl) or 50 nM DBCO-tagged Cy3-labeled ssDNA (in 50 mM pH 7.4 sodium phosphate buffer, 100 mM NaCl, which also contained 1 mM $Mg^{2+}$) was carefully cast on the functionalized surface of each diamond and incubated at room temperature in a dark, moisturized environment for 20 min. For the streptavidin-mediated system, 20 nM biotin-tagged Cy3-labeled ssDNA was premixed with 40 nM non-labeled streptavidin at 1:1 volume ratio and incubated at room temperature for 20 min with mild agitation, which should predominantly result in 1-ssDNA:1-streptavidin conjugates whose effective concentration was 10 nM. This solution was then applied to diamond surfaces in the same way. Upon completion, the diamond was gently rinsed with the same buffer and placed in an imaging dish for fluorescent microscopy study. The primary sequence of the 40-nt ssDNA is 5'-TTTTT TTTTT AGTCC GTGGT AGGGC AGGTT GGGGT GACTT-3'. The biotin-tag (or DBCO-tag), followed by a Cy3-label, is attached to the 5' end of the ssDNA.

### Single-molecule imaging

Fluorescent single-molecule imaging was performed on a custom-built fluorescence microscope equipped with 488 nm and 532 nm lasers (Coherent Sapphire) and a 60X oil TIRF objective (Olympus UPLAPO60XOHR) in an inverted configuration. Diamond samples were placed inside a buffer-containing dish with the functionalized side facing down, and imaged through the glass-coverslip bottom that was also passivated with mPEG to minimize non-specific binding (see Fig. S5 for an illustration). For 488 nm excitation (SA-488), a ZET488/10x (Chroma) notch filter, a ZT488rdc-UF1 dichroic beamsplitter, and an ET525/50m emission filter were used. For 532 excitation (Cy3-ssDNA), a ZET532/10x notch filter, a ZT532rdc-UF1 dichroic beamsplitter, and an ET575/50m emission filter were used. Images were acquired by an Andor iXon Ultra 888 EMCCD camera (EMCCD cooled down to −60°C) with 1 s exposure time and 200 (or 300) gain, or 500 ms × 120 for video.

### Resetting surface

Diamonds were first soaked in 1 M KOH (typically overnight but can be shortened), which effectively removes $Al_2O_3$ at a rate of 3.6 nm/min (Fig. S6). They were then immersed in NanoStrip (99% sulfuric acid + 1% $H_2O_2$) at room temperature for 5 min, rinsed extensively with water, sonicated in acetone and 2-isopropanol for 5 min each, and finally dried with nitrogen gas. This cleaning procedure proved to be reliable to restore the diamond for single-molecule fluorescent imaging experiments. For spin coherent measurements, while the above-mentioned regenerating procedure may be sufficient, we in practice replaced the NanoStrip treatment by tri-acid cleaning, which uses a 1:1:1 mixture of nitric acid, perchloric acid, and sulfuric acid at boiling temperatures (34) to ensure minimal contamination.

**Coherence measurement**
Coherence measurements were performed at 1750 G magnetic field strength. We use both YY-8 and spin-echo pulse sequences to measure the coherence and depth of NV centers. For YY-8, $F_0(t)$ and $F_1(t)$ measure the projection of the NV spin coherence on $|0\rangle$ and $|-1\rangle$ states, respectively, depending on the phase of the last $\pi/2$ pulse. The normalized signal (spin contrast) is defined by $2[F_0(t) - F_1(t)] / [F_0(t) + F_1(t)]$, which removes the common-mode noise. The coherence data are fitted to a stretched exponential function $\exp[-(t/T_2)^n]$. $T_2$ and number of $\pi$-pulse data are fitted to either a saturation curve

$$T_2(N) = T_2(1)[N_{sat}^s + (N^s - N_{sat}^s) \exp(-N/N_{sat})] \quad , \qquad [1]$$

or to a power law

$$T_2(N) = T_2(1) N^s \qquad [2]$$

when there is no observable saturation of $T_2$ (15). The final fitting parameters are given in Table S2.

**Estimated integration time for single $^{13}$C spin detection**
Based on the experimental parameters in Figure 4B we estimate the required integration time to detect a target nuclear spin. In NV sensing, the signal from an individual target spin is given by Signal=$\frac{T}{\tau+t_{read}} \frac{f_0-f_1}{2} \chi(\tau) |\sin(A\,\tau/\hbar)|$, where $T$ denotes the total measurement time, $\tau$ the phase accumulation time, $t_{read}$ the optical spin-readout time, $f_0 = 0.063$ ($f_1 = 0.048$) the average number of detected photons per readout window in $m_s = 0$ ($m_s = 1$), $\chi$ the spin contrast from Fig. 4B, $A$ the hyperfine coupling between the NV and the target spin, and $\hbar$ the reduced Planck constant. The noise (standard deviation) is given by Noise=$\sqrt{\frac{T}{\tau+t_{read}} \frac{f_0+f_1}{2}}$. The NV center in Figure 4B has a depth of 4.8 nm, with an additional 5 nm for the surface functionalization. Taking into account that the diamond has a (1,0,0) cut, we estimate the average interaction strength between NV centers and $^{13}$C spins to be $A= (2\pi)160$ Hz. Based on these parameters, we estimate the required integration time to be 2.8 hours. These demanding integration times can further be reduced to 100 seconds by utilizing quantum logic sequences (e.g. 100 repetitions) (1).

**Acknowledgments**

We thank Drs. Nazar Delegan and Joseph Heremans for insightful discussions on diamond surface termination; Xinghan Guo, Dr. Noelia Bocchio and Prof. Alexander High for help on ALD deposition and patterning; Dr. Jon Monserud, Sheela Waugh, Dr. Jason Cleveland, and Larry Gold for providing Cy3-ssDNA constructs and discussions on single-molecule imaging; and Uri Zvi, Zhendong Zhang, Dr. Joonhee Choi and Hengyun Zhou for discussions and insights on quantum sensing. M.X., X.Y. and P.C.M. acknowledge financial support from the National Science Foundation (NSF) Grant No. OMA-1936118, OIA-2040520 and the Swiss National Science Foundation (SNSF) Grant No. 176875. N.Q. and A.T. acknowledge financial support from the SNSF Grant No. 183717. L.V.H.R. acknowledges support from the DOD through the NDSEG Fellowship Program. N.P.d.L. acknowledges support from the National Science Foundation (NSF) Grant No. OMA-1936118 and DMR1752047. The authors acknowledge the use of the Pritzker Nanofabrication Facility at the University of Chicago (NSF ECCS-2025633), the University of Chicago Materials Research Science and Engineering Center (DMR-2011854), as well as the Imaging and Analysis Center at Princeton University (DMR-2011750).

# Figures and Tables

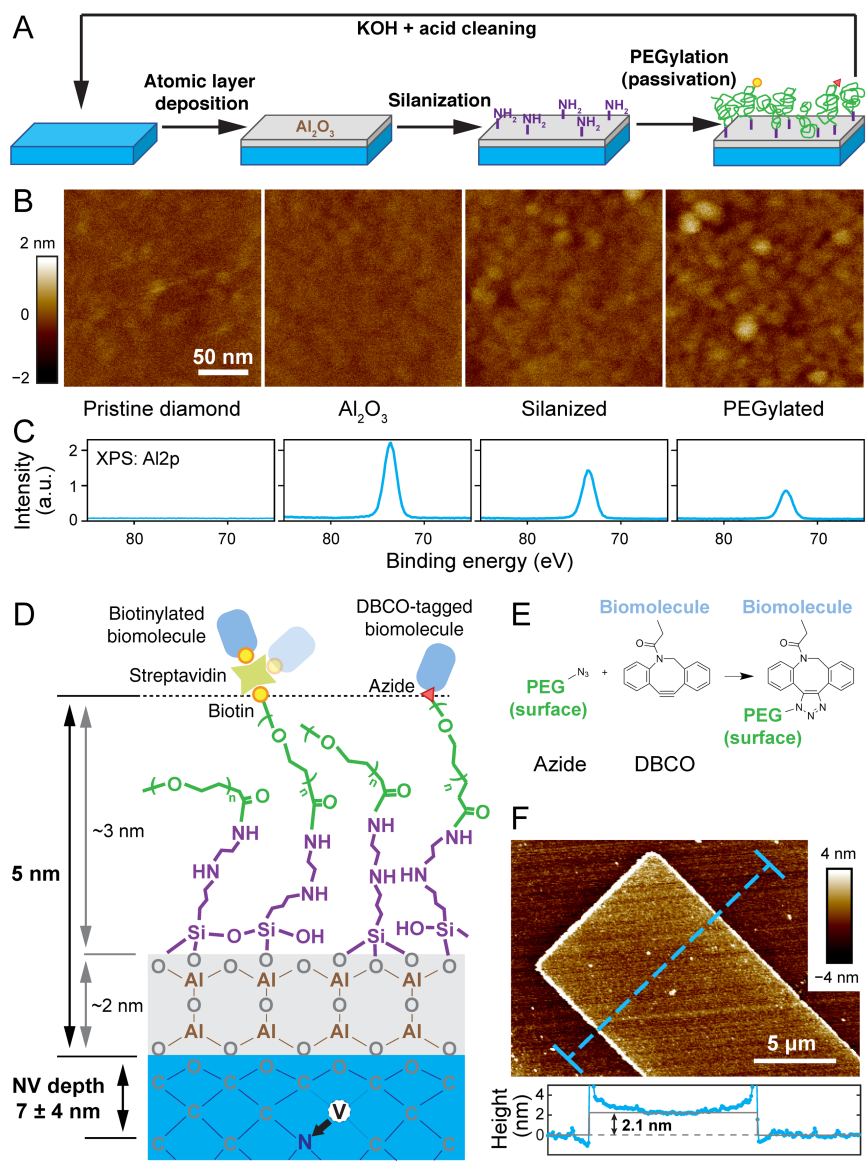

**Figure 1.** Architecture and characterization of the diamond functionalization approach. (*A*) Schematic illustration of the functionalization process. A thin layer of $Al_2O_3$ (gray) was deposited to the pristine oxygen terminated diamond surfaces (blue), followed by silanization (purple) and PEGylation (green). Functional groups (biotin, yellow circle; azide, red triangle) allow for crosslinking with target biomolecules. (*B*) AFM characterization of the surfaces and (*C*) XPS Al2p signal after each step of the functionalization. (*D*) Illustration of the overall chemical functionalization architecture (not to scale), with corresponding thicknesses. (*E*) Illustration of SPAAC reaction. (*F*) A lithographically fabricated $Al_2O_3$ pattern on the diamond surface by lift-off, with a thickness of approximately 2.1 nm. The $Al_2O_3$ layer is uniform without the presence of pin holes. The elevated edges originate from lift-off combined with ALD deposition.

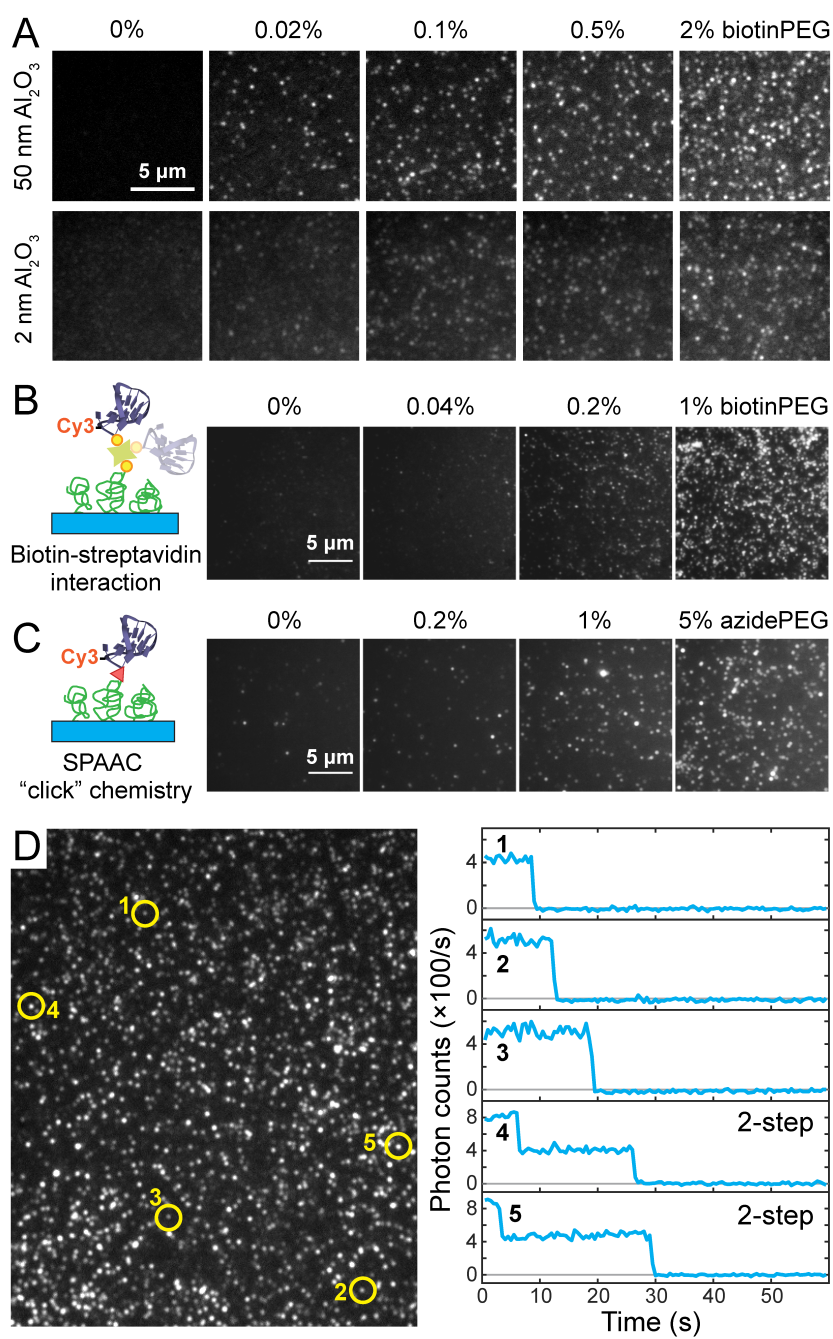

**Figure 2.** Single-molecule characterization of fluorescently labeled biomolecules immobilized on the diamond surfaces. (*A*) Fluorescence images of the immobilized SA-488 molecules for various biotinPEG percentages (0%, 0.02%, 0.1%, 0.5%, and 2%) and two different $Al_2O_3$ thickness (50 nm, imaged in buffer; and 2 nm, imaged in a refraction index = 1.42 anti-fade medium). (*B*, *C*) Immobilization of a Cy3-ssDNA on diamond surfaces. This is achieved via either (*B*) biotin-SA interactions or (*C*) SPAAC. (*D*) A representative area of single-molecule fluorescent images of (*B*) and the time traces of 5 selected fluorescent spots.

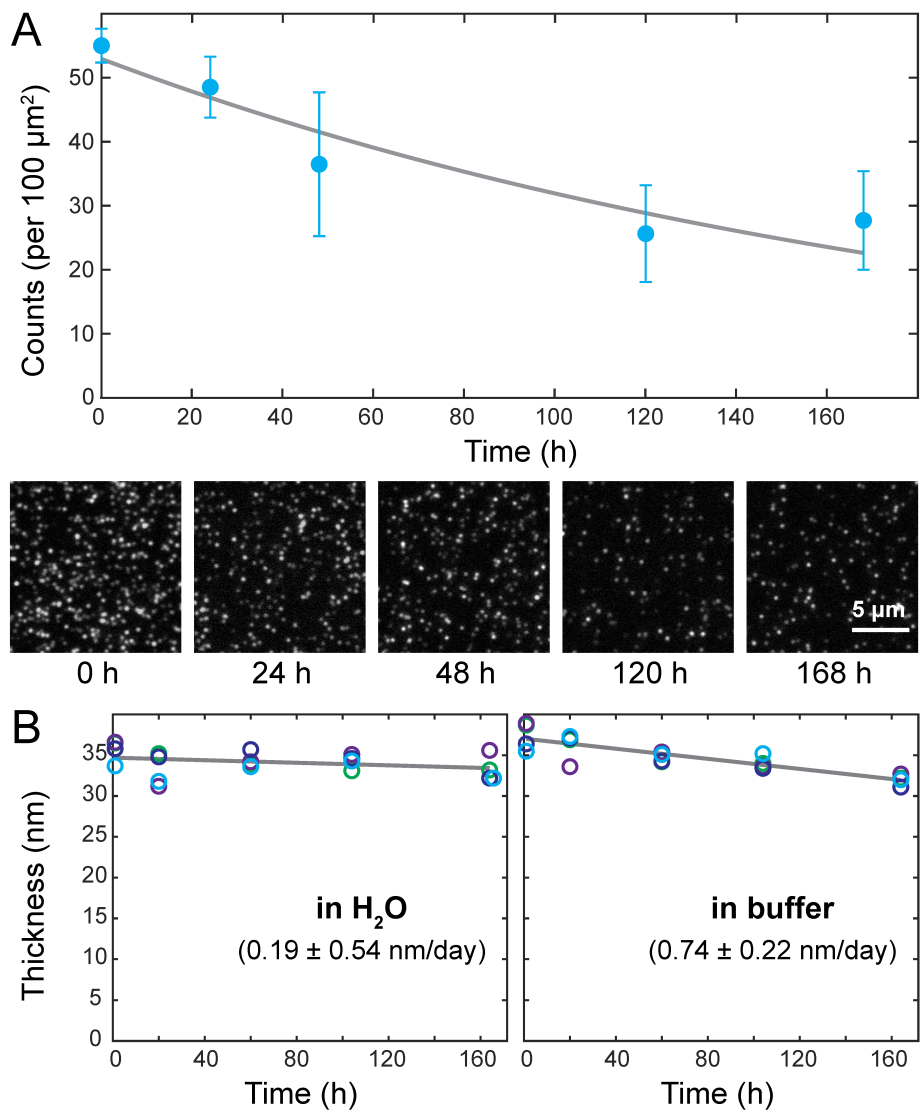

**Figure 3**. Time stability of the functionalization architecture in a physiological relevant environment. (*A*) Number of SA-488 molecules per 100 µm² area (blue circles) detected by single-molecule fluorescence microscopy as a function of storage time in sodium phosphate buffer (pH 7.4, [$NaH_2PO_4$ + $Na_2HPO_4$] = 50 mM, [NaCl] = 100 mM) over a course of one week. Each data point is based on three 2800 µm² field-of-view areas; error bars indicate one standard deviation. Fit is an exponential decay. Representative single-molecule microscopy images on a 50 nm thick $Al_2O_3$ layer are displayed at the bottom. (*B*) The overall thicknesses of the functional layer prepared on a 35 nm-thick lithographically patterned $Al_2O_3$ structure in $H_2O$ (left) and sodium phosphate buffer (right) were tracked by AFM over a course of one week at room temperature. Four unique sites (circles of the same color) were monitored for each sample, and the mean values were fitted to a linear model (gray), which allowed us to estimate the dissolution rates.

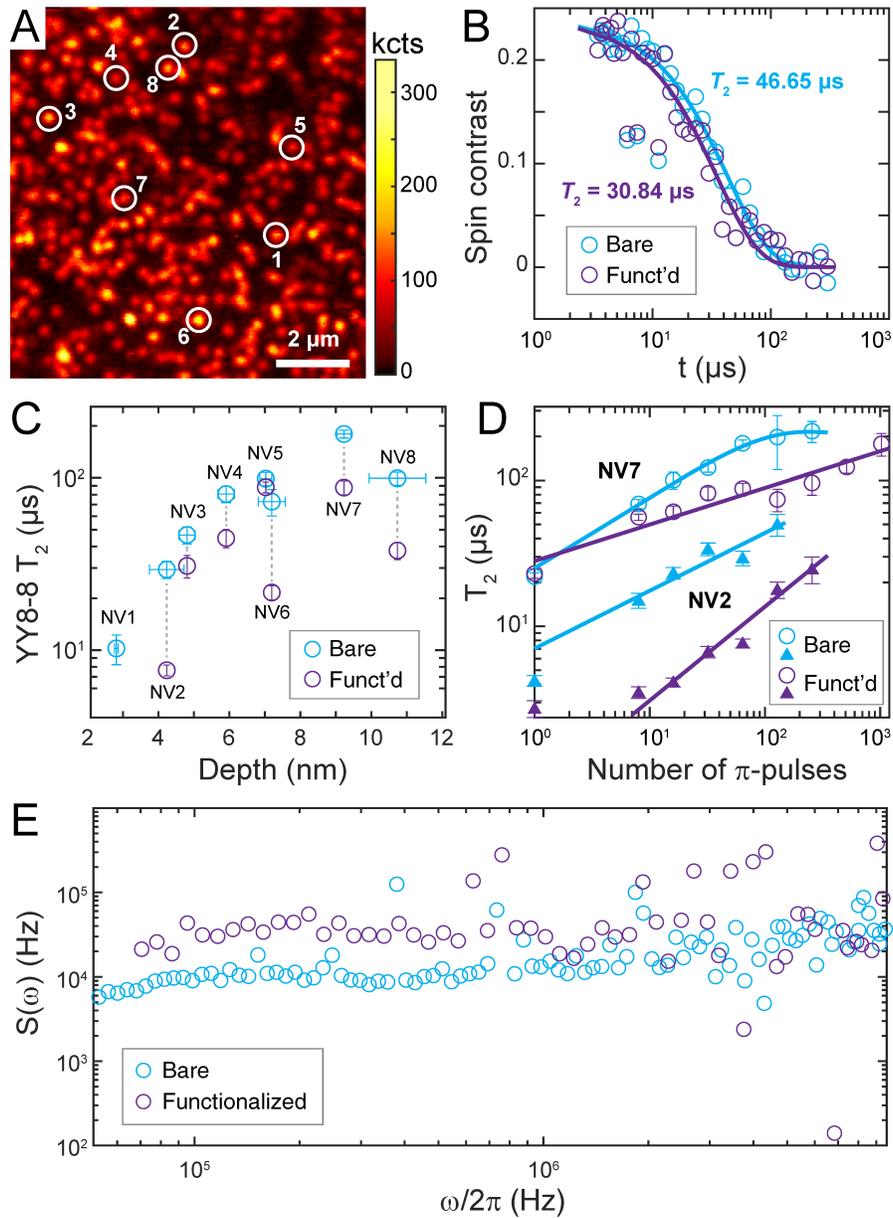

**Figure 4**. Impact of functionalization on NV electron spin coherence. (*A*) Confocal scan of near surface NV centers (implantation energy 3 keV). The eight NV centers studied for their coherence times are marked by circles. (*B*) Typical time trace of coherence measured by a (YY-8)$_{N=8}$ sequences before (blue) and after (purple) functionalization for NV center number 3 (depth 4.8nm). $T_2$ times are based on the fitted stretched exponential decays (solid lines). (*C*) $T_2$ measured by (YY-8)$_{N=8}$ pulse sequence (total of 64 $\pi$-pulses) plotted against NV depth before (blue) and after (purple) functionalization. Depth calibration was performed following Ref. (23). (*D*) $T_2$ times as a function of number of $\pi$-pulses for NV number 2 (depth 4.2 nm, triangles) and NV number 7 (depth 9.2 nm, open circles) before (blue) and after (purple) functionalization. Solid lines are fits based on equation (1) and (2). (*E*) Spectral decomposition manifests a broadband noise spectrum across the frequency range of 0.05-10 Mhz for NV number 7. All measurements were carried out at 1750 G magnetic field strength.